**Estimating quantile treatment effect on the original scale of the outcome variable: a case study of common cold treatments**

Harri Hemilä [1] and Matti Pirinen [2]

[1] Department of Public Health, University of Helsinki, FINLAND

[2] Institute for Molecular Medicine Finland (FIMM), HiLIFE, Department of Public Health, and

Department of Mathematics and Statistics, University of Helsinki, Helsinki, Finland.

Corresponding author:
Harri Hemilä, MD, PhD
Department of Public Health,
University of Helsinki, POB 20,
Helsinki, FI-00014, FINLAND.
harri.hemila@helsinki.fi
https://orcid.org/0000-0002-4710-307X

Matti Pirinen, PhD
https://orcid.org/0000-0002-1664-1350

**Words 4014  --- 2023-10-22**

**Abstract**

The effects of treatments on continuous outcomes can be estimated by the mean difference (i.e. by measurement units) and the relative effect scales (i.e. by percentages), both of which provide only a single effect size estimate over the study population. Quantile treatment effect (QTE) analysis is more informative as it describes the effect of the treatment across the whole population. A drawback of QTE has been that it is usually presented over the quantiles of the control group distribution, whereas presentation over the measurement units is often more informative.

We developed a method to estimate back-transformed QTE (BQTE), that presents QTE as a function of the outcome value in the control group, using piecewise linear interpolation and bootstrapping. We further applied the BQTE function to provide informative bounds on the treatment effect at the upper and lower tails of the population. To illustrate the approach, we used 3 data sets of treatment for the common cold: zinc gluconate lozenges, zinc acetate lozenges, and nasal carrageenan. In all data sets, the relative scale provided a better summary of the BQTE distribution than the mean difference.

The BQTE approach is particularly useful for describing the variability of effects on the duration of illnesses, length of hospital stay and other continuous outcomes that can vary greatly in the population. Using this method, it is possible to present the QTE by the measurement units, which provides an informative addition to the standard presentation by quantiles.

**Introduction**

In meta-analyses of continuous outcomes, estimating the treatment effect on the standardized mean difference (SMD) scale has been very common [1]. In the SMD approach, the difference between treatment and control groups is normalized to the standard deviation (SD) scale so that in each study, one unit indicates the SD of that trial. However, while the SMD approach allows for analysis of a combination of studies with different measurement scales, it may be difficult to interpret the pooled findings. For example, few medically-oriented readers can form an informed opinion as to whether a 0.97 SD unit effect of zinc on common cold duration [2] is large or small. Unsurprisingly, a recent survey of physicians concluded that "presenting results as a standardized mean difference, the longest standing and most widely used approach, was poorly understood and perceived as least useful" [3].

Another popular scale to present continuous outcomes is the mean difference (MD) scale, which is based on the units of the empirical observations [1]. For a reader, the effect on the MD scale may appear easy to interpret, such as colds which are, on average, 1.03 days shorter in the zinc groups [4]. However, many continuous outcomes have wide distributions. For example, the duration of the common cold can vary from less than 1 day to over 2 weeks. The 1-day colds cannot be shortened by over 1 day and therefore the above-described average treatment effect (ATE) estimate on the MD scale cannot be a meaningful description of the difference at the lower tail of the distribution.

A third approach for estimating the effect on continuous outcomes is to use the relative scale, calculating the ratio of means (RoM) [5]. In this approach, the control group mean is the unit of the measurement scale, such that the RoM between the treatment and control groups gives the effect estimate. The relative scale has been very common over decades in the analysis of binary outcomes, but it has been rare in the analysis of continuous outcomes. Nevertheless, there are strong arguments that many continuous outcomes, such as the duration of illness, should also be analyzed on the relative scale since that approach adjusts for some baseline variation between different trials, and appears to better capture the effect within trials [5,6]. The RoM estimates can be presented as percentages, which are familiar to most people.

All three approaches described above use a single value to describe the average effect of the medical intervention over the population. A fourth approach, the quantile treatment effect (QTE), describes the effect of the treatment by comparing the entire outcome distributions between the treatment and control groups [7-11]. In the QTE approach, the difference between the treatment and control groups is shown on the vertical axis, with the horizontal axis showing the quantiles of the outcome distribution in the control group. The QTE function describes the effect of the treatment on the population in more detail than the three average effects listed above. Furthermore, with an additional assumption that the treatment preserves the relative order of the outcome value among the treated compared to the counterfactual case [12] where they were not treated, QTE gives an estimate of the treatment effect for each quantile level of the untreated population. Nevertheless, a shortcoming of the QTE approach has been that the quantile levels in the untreated control group are not as intuitive as the units of the original outcome variable. For example, in the Mossad et al. (1996) trial [13], the QTE of zinc gluconate lozenges was 8 days at the $80^{th}$ percentile [11], but it remains unclear what control group duration corresponds to this reduction of 8 days.

In this paper we propose an approach to estimate the back-transformed quantile treatment effect (BQTE) function together with an estimate of its uncertainty. BQTE presents QTE on the empirical outcome scale so that the interpretation of the estimates is more evident. First, we use bootstrapping to generate a set of paired distributions for the treatment and control groups. Second, we fit a piecewise linear approximation of the BQTE function to each bootstrapped set of paired distributions. Finally, we calculate the distributions of both the BQTE and its ratio to the control group value as functions of the outcome value in the control group. This approach also allows a comparison of whether the ATE scale or the RoM scale better captures the observed effect of the treatment on a specific data set. We explain the connection between BQTE and the theoretical concept of individual-level treatment effect and we show how the former can provide informative bounds on the latter at the tails of the outcome distribution. We verify our methodology using simulations and illustrate the approach on the data from a zinc gluconate lozenge trial [13], pooled data from three zinc acetate lozenge trials [14-16], and pooled data from two nasal iota-carrageenan trials [17,18].

**Methods and Materials**

Estimation of BQTE and its Uncertainty

Let $F$ and $G$ be the cumulative distribution functions of the control and treatment populations, respectively. For any quantile level $p \in (0,1)$, the quantile treatment effect is defined [8-10] as

$$\text{QTE}(p) = G^{-1}(p) - F^{-1}(p).$$

QTE describes the effect of treatment by comparing the outcome values in the treatment and control populations as a function of quantile levels. We, instead, aim to present this information as a function of the outcome value itself. For that, we transform QTE on the outcome scale by defining a back-transformed quantile treatment effect function at outcome value $x$ in the control population as

$$\text{BQTE}(x) = \text{QTE}(F(x)) = G^{-1}(F(x)) - x.$$

While this quantity has been considered before [7], it does not seem to be commonly applied even though it would provide an informative extension of the standard QTE presentation. Next, we explain our method to estimate $\text{BQTE}(x)$ at any given point $x$ together with the uncertainty related to the estimate. We have implemented the method in an R function called 'bqte' which is freely available at [19].

Suppose that we have samples of observations from the treatment and control populations with possibly different sample sizes. To estimate the BQTE function, we simultaneously generate bootstrap samples of the outcome values in the treatment and control groups. For each paired bootstrap sample, we calculate the empirical quantiles of both groups at $K$ cut-points, corresponding to the quantile levels of $1/(K+1), \ldots, K/(K+1)$, where $K$ is the sample size of the control group. Denote these quantiles by $x_i$ in the control group and $y_i$ in the treatment group, for $i = 1, \ldots, K$. We now have estimates of QTE and BQTE functions at $K$ points:

$$\widehat{\mathrm{BQTE}}(x_i) = \widehat{\mathrm{QTE}}(i/(K+1)) = y_i - x_i.$$

We then apply linear functions to estimate BQTE between these $K$ values as

$$\widehat{\mathrm{BQTE}}(x) = y_i + \frac{x - x_i}{x_{i+1} - x_i}(y_{i+1} - y_i) - x, \text{ when } x \in (x_i, x_{i+1}).$$

By default, we use the mean of the bootstrap distribution of $\widehat{\mathrm{BQTE}}(x)$ as the final estimate of BQTE because that approach reduced the mean-square error (MSE) of the estimator compared to a direct application of $\widehat{\mathrm{BQTE}}(x)$ on the observed data set in our simulation study. Our implementation also allows the direct estimator of BQTE without the bootstrap averaging. We define level α confidence intervals for BQTE by empirical quantiles of the bootstrap distribution corresponding to probability values of α/2 and 1 – (α/2). We restricted the estimation of BQTE between the quantiles of the control group corresponding to probability values of 5/$K$ and 1 – (5/$K$). This range was chosen based on our simulation study assessing the validity of our approach.

Connection between BQTE Distribution and Individual-level Treatment Effect

The BQTE function is a population-level measure of the effect of the treatment. We could also ask how the treatment would have affected a particular individual whose outcome value without the treatment is known. As we explain in Supplementary Information, such an individual-level treatment effect can be estimated by the BQTE function only when we assume that the treatment preserves the order of outcome values in the study population. However, even without such an assumption, the BQTE function can be used to provide upper and lower bounds for the ATE at the tails of the control distribution. We define the upper tail back-transformed quantile treatment effect (UTBQTE) and the lower tail back-transformed quantile treatment effect (LTBQTE) for any outcome value $x$ in the control population as differences in the average outcome values between the corresponding upper and lower tails of the treatment population ($Y$) and the control population ($X$):

$$\text{UTBQTE}(x) = E\big(Y \mid Y \geq G^{-1}(F(x))\big) - E(X \mid X \geq x),$$

$$\text{LTBQTE}(x) = E\big(Y \mid Y \leq G^{-1}(F(x))\big) - E(X \mid X \leq x).$$

We show in Supplementary Information that UTBQTE gives an upper bound for the ATE of the upper tail and LTBQTE gives a lower bound for the ATE of the lower tail. Our freely-available implementation of the BQTE function includes also the computation of these bounds together with their confidence intervals estimated by the bootstrap method.

Simulation Study

We validated our methodology and implementation using a simulation study reported in detail in Supplementary Information. First, we assessed which were the suitable quantile levels as a function of sample sizes where our estimator performed reliably, that is, bias was small and the confidence intervals were well-calibrated. Second, we assessed how the number $K$ of the cut points used in the piecewise linear interpolation affected the results. Finally, we compared two versions of our estimator for the BQTE function, namely the bootstrap averaging (bagging) and the direct estimator, with the previously introduced Doksum's estimator [7].

Statistical Analysis of Common Cold Trials

We report BQTE on both direct and relative scales. The direct scale reports BQTE as the difference in the disease duration between the groups, which gives the effect of zinc lozenges and nasal carrageenan on days and corresponds to the ATE approach. The relative scale reports the ratio between BQTE and the disease duration in the control group, corresponding to the RoM approach.

In our figures, we present the relative BQTE in percentages. We show the 95% confidence intervals (CI) for all estimates. We used 2,000 bootstrap samples.

Code Availability

The R functions and the example code for the calculation of the Mossad et al. trial are available at GitHub [19].

Data of the Included Trials

The contexts of the zinc lozenge trials, and the characteristics of the patients are described in the trial reports [14-16], and summarized in previous analyses [6,20,21]. In brief, all trials were randomized, double-blind and placebo controlled. Mossad et al. (1996) studied employees of the Cleveland Clinic (mean age 38 y) [13]. The first zinc acetate lozenge trial recruited volunteers from the University of Texas (mean 26 y) [14], and the two other trials recruited volunteers from Detroit Medical Center (mean 35-37 y) [15,16]. All trials recruited patients with natural common colds acquired in the community, and all trials tested the treatment effect of zinc lozenges.

Mossad et al. (1996) published the findings of their trial on zinc gluconate lozenges as survival curves [13]. The numbers of common cold patients recovering each day were measured from the curves [21]. The data sets of the three randomized trials on zinc acetate lozenges (Petrus et al., 1998; Prasad et al., 2000; Prasad et al., 2008) [14-16] were provided by the authors of the trials and were used in previous individual patient data meta-analyses of common cold duration [20] and recovery rate [21]. In our meta-analysis of the zinc acetate lozenge trials, we combined the three data sets. The zinc lozenge data sets are available [19-21].

In the Mossad trial, there were eight censored observations (8% of the total) [13]. In the placebo group there were two on day 7, one on day 15, one on day 16, and two on day 19; and in the zinc treatment group there was one on day 9 and one on day 11. We imputed the duration as the day of censoring. Four of the censored observations were beyond the 93rd percentile and this imputation has minimal influence on our analysis. There were no censored observations in the zinc acetate lozenge trials [14-16].

The characteristics of the patients and the contexts of the two nasal iota-carrageenan trials are described in the trial reports [17,18], and were summarized in previous meta-analyses [22,23]. In brief, the Ludvig et al. (2013) trial with adults (mean age 33 y) [17] and the Fazekas et al. (2012) trial with children (mean 5 y) [18] were randomized, double-blind, placebo-controlled trials, carried out in Vienna, Austria. For enrollment, participants were required to have mild to moderate common cold symptoms of natural origin. The numbers of common cold patients recovering each day were measured from the curves [23]. In these two trials, there were 6 censored observations in the carrageenan groups and 21 in the placebo groups, all on day 20. We imputed the duration as the day of censoring. In our bootstrap approach, right-hand censoring in the control group typically has little influence on the estimation of BQTE on days prior to censoring. In a sensitivity analysis, we imputed 30 days for the 27 censored observations, but observed no notable difference from our original results (see Supporting Information Fig. S8). The carrageenan data set is available [19,23].

**Results**

Figure 1 demonstrates our approach to estimate the BQTE function and its uncertainty using 20 bootstrap samples of the Mossad et al. trial [13].

Our simulation study results are reported in Supplementary Information. As a summary, we observed that we need to restrict the estimation of BQTE and QTE functions between the quantile levels of $5/N$ and $1-5/N$, where $N$ is the sample size in the control group in order to have reliable results (Fig. S2). Our default choice for the number of cutoff points is $K = N$ based on the results reported in Fig. S3. In Fig. S4, we compared the root-mean-square-error (RMSE) of Doksum's estimator and the two versions of our estimator and found smaller RMSE for our estimators. Additionally, we observed that the bagging estimator performed better than the direct estimator and therefore we use the bagging estimator as our default choice.

The analysis of the zinc gluconate lozenges in the Mossad et al. trial [13] is shown in Figure 2. The red dots indicate the estimated BQTE for each value of the disease duration in the control group and the vertical lines indicate the corresponding 95% CI. The estimation was restricted between the 1$^{st}$ and 9$^{th}$ deciles of the control group, corresponding to 3 and 17 days, respectively, in order to have sufficient information about the accuracy of the BQTE estimator.

Figure 2A shows BQTE in days, corresponding to the ATE scale. The overall ATE of zinc gluconate lozenges is a 4.0-day reduction in common cold duration, which is indicated by the horizontal dotted line. The 95% CIs for the BQTE function before the 7$^{th}$ day or after the 11$^{th}$ day do not overlap with ATE. In addition, the 95% CI of the BQTE estimates for the longest durations are far apart from the 95% CIs of the estimates for the shortest durations, indicating that the BQTE estimates at the two extremes of the distribution are very different from each other when measured in days. If we further assume that the treatment preserves the ranking of the colds, then this analysis would suggest that untreated colds that last for 3 days are expected to be shortened on average by about 1 day, whereas 15-17 day colds are expected to be shortened by about 8 days. Even without any assumptions about whether the treatment preserves the ranking, we can conclude that the average shortening among the untreated colds of 15 days or more is at least 5.7 days, as the 95% CI of UTBQTE at 15 days is from 5.7 to 9.8 days (Supplementary Information Fig. S5).

The relative BQTE of the zinc lozenges in the Mossad et al. trial is shown in Figure 2B. The dotted horizontal line indicates the overall relative effect of a 43% reduction in common cold duration, corresponding to RoM = 0.57. The 95% CIs throughout the distribution are consistent with a constant relative BQTE, with all the estimated relative BQTEs located between 30% and 50%. Thus, the average reduction of 43% in common cold duration seems to adequately describe the BQTE throughout the distribution, including both the shortest and longest colds (Fig. S5).

Figure 3 shows an analysis of pooled results of the three zinc acetate lozenge trials [14-16]. Again, BQTE function of zinc lozenges on the day scale is not described well by the overall ATE of a 2.7-day reduction (Fig. 3A). BQTE was substantially smaller at disease duration of 5 days or less in the untreated group. However, the 95% CIs at the long colds are wide and overlap with ATE. On the relative scale, the mean effect of 36% seems to describe better the whole BQTE distribution, with some potential deviation towards smaller relative BQTE values observed at the shortest durations (Fig. 3B; Supplementary Information Fig. S6).

Figure 4 shows a similar analysis of the pooled results of the two nasal iota-carrageenan trials [17,18]. The BQTE function of nasal carrageenan on the day scale is not described well by the estimated overall ATE of a 1.9-day reduction (Fig. 4A). The 95% CIs for the BQTE of carrageenan at the shortest and longest colds do not overlap on the day scale. In this case, the relative BQTE is not as uniform as in the zinc lozenge trials. Instead, the relative BQTE appears to be greater for longer colds, although the 95% CIs are so wide that no definite conclusion can be drawn (Fig. 4B).

There is no indication of a meaningful benefit for patients who have colds shorter than one week as both BQTE and LTBQTE remain zero in this region (Fig. 4 and Supplementary Information Fig. S7). However, for patients who would have colds lasting two weeks or more, based on the BQTE estimate, carrageenan may shorten colds by up to 5 to 7 days (Fig. 4A) and, based on the UTBQTE estimate, at least by 1.8 days on average (Fig. S7). On the relative scale, the two-week colds could be shortened by up to 30% to 40% (Fig. 4B) and, based on the UTBQTE estimate, the average effect on the two-week colds is at least 6% (Fig. S7).

**Discussion**

Quantile regression is a well-established statistical method, which has been widely used in econometrics [24-30]. So far, it has been little used in clinical medicine though its use has been encouraged by several authors [31-35].

The QTE approach is a special case of quantile regression and can be particularly useful in the analysis of randomized trials and other controlled studies since it enables investigation of how the treatment affects the whole distribution of a continuous outcome of untreated patients [7-11,23,36-38]. The QTE approach can be used to examine treatment effects on the duration of various illnesses, the duration of hospital stay and ICU stay, and on comparable continuous outcomes. These outcomes usually have asymmetric distributions with fat tails, and censored observations at the right-hand tail are not uncommon, yet QTE analysis allows for all these features.

QTE values are usually shown on the vertical axis as a function of the quantiles of the outcome in the untreated population on the horizontal axis [8-11]. An attractive interpretation of QTE is as the average benefit from the treatment for patients in a particular quantile of the untreated distribution, for example, in the first decile or in the last decile. Conceptually, however, this interpretation requires that the treatment preserves the relative order of the outcome value among the treated compared to the hypothetical case where they were not treated [12] (See Supplementary Information for a detailed discussion). Since there is no way to confirm that the treatment operates in this way, strictly speaking, QTE is a quantity related to populations rather than to any single individual. The same is true for our BQTE values that transform the QTE function back to the original observation scale. However, using the BQTE function, we are able to derive general upper and lower bounds on the average treatment effect of the individuals from the upper or lower tails of the untreated population without making any assumptions about how the treatment operates (see Supplementary Information). This information provides an informative extension of the standard

ATE approach that uses a single value to describe the average treatment effect on the whole population.

In medicine, often a clinically more relevant question is how large the effect is on patients with a particular duration of disease rather than on patients in a particular quantile of the disease duration. The impact of treatment on a disease of long duration is often much more relevant than shortening an already short disease. In calculating the overall ATE, the latter may mask substantially larger effects on the former. By comparing the entire distributions of disease durations, our approach captures more information about the effect of the treatment than a single estimate of ATE. It can be applied to single RCTs as in Fig. 2, and to individual-patient data (IPD) meta-analyses as in Figs. 3 and 4. Our approach applies on IPD but even when only a limited number of controlled studies included in a meta-analysis have IPD available, our approach can be used to test whether the RoM or the ATE approach gives an estimate that better describes the full BQTE distribution. Thereafter the study-level findings can be combined using the more appropriate scale.

The sample size of the study affects the range of quantiles at which our bootstrap approach provides valid confidence intervals. Based on a simulation experiment reported in Supplementary Information Figure S2, we suggest applying the approach between the quantiles $5/K$ and $1 - (5/K)$ where $K$ is the sample size in the control group. For example, in the Mossad et al. trial [13], $K = 50$ and thus our proposed range of BQTE estimation extended from the 1$^{st}$ to the 9$^{th}$ decile of the control group values which corresponded to 3 and 17 days, respectively.

Conceptually, our analyses are important in illustrating that the ATE estimates, that are widely used in meta-analyses [1], appear to seldom describe properly the whole BQTE distribution. In all three data sets, there was a strong discrepancy between the ATE estimates and the BQTE estimates at the extreme ends of the outcome distribution of the untreated group (Figs. 2A, 3A and 4A). As a separate approach, we used the UTBQTE and LTBQTE functions to calculate bounds for the treatment effects in the tails, and we showed that the ATEs at the tails deviate from the overall ATE estimate no matter how the treatment operates, i.e. without any assumptions about order preservation between the control and treatment groups (Supplementary Information Fig. S5, S6 and S7). In general, the discrepancy between the BQTE function and the ATE estimate is the smallest when the treatment preserves order whereas the variance of the treatment effect increases when the treatment deviates more from maintaining the order (Supplementary Information).

In both zinc lozenge data sets, the relative scale estimates of BQTE were consistent with a constant value throughout the distribution (Figs. 2B and 3B). This gives further support for a wider use of the relative scale in the analysis of continuous outcomes [5,6,39]. However, our findings should not be interpreted as an indication that in all cases the relative BQTE is necessarily constant over the whole range of the outcome value in the control group. Rather, our approach provides an easy tool to assess whether the relative BQTE is constant over the control outcome distribution in a particular case. The BQTE of carrageenan on the relative scale does not appear to be constant but rather seems greater at long colds, although the confidence intervals for the effects are wide and a firm conclusion cannot be drawn (Fig. 4B). Our comparisons of the BQTE approach with the ATE and RoM estimates suggests that often the latter is preferable.

We do not propose that our approach should replace the standard plotting of QTEs by the quantiles of the outcome distribution in the control group. Both ways of presenting data have their strengths, and it might be most informative to present results by both approaches.

**Acknowledgements**
We are grateful to Elizabeth Chalker for critically reading the manuscript.

**Author contributions**
**Harri Hemilä:** conceptualization, methodology, investigation, writing original manuscript.
**Matti Pirinen:** conceptualization, methodology, investigation, review and editing of the manuscript.

**Conflicts of interest**
None.

**Funding**
No external funding.

**Availability of Data and Materials**
The datasets generated and/or analysed during the current study are available [19]:
https://github.com/mjpirinen/bqte

**Supporting Information**
File format: .pdf.
Contents: Interpretation of BQTE and testing the implementation of the bqte( ) function.
Available at [19]:
https://github.com/mjpirinen/bqte

**ORCID**
Harri Hemilä
https://orcid.org/0000-0002-4710-307X

Matti Pirinen
https://orcid.org/0000-0002-1664-1350

**Figure legends**

**Fig. 1.** Demonstration of the proposed method to estimate back-transformed quantile treatment effect (BQTE) as a function of outcome value in the control group. There are $B = 20$ piecewise linear estimates of the BQTE function, each based on one pair of bootstrap samples of the treatment and control groups of the Mossad et al. data [13]. For any given value on the x-axis (duration of cold in the control group), we estimate the BQTE as the mean of the corresponding y-axis values over the $B$ bootstrap samples. The red dot is BQTE estimate at $x = 10$ days. A level $\alpha$ confidence interval (CI) for BQTE is defined by quantiles of level $\alpha/2$ and $1 - (\alpha/2)$ of the bootstrap sample (e.g. red segment is 80% CI at $x = 10$ days).

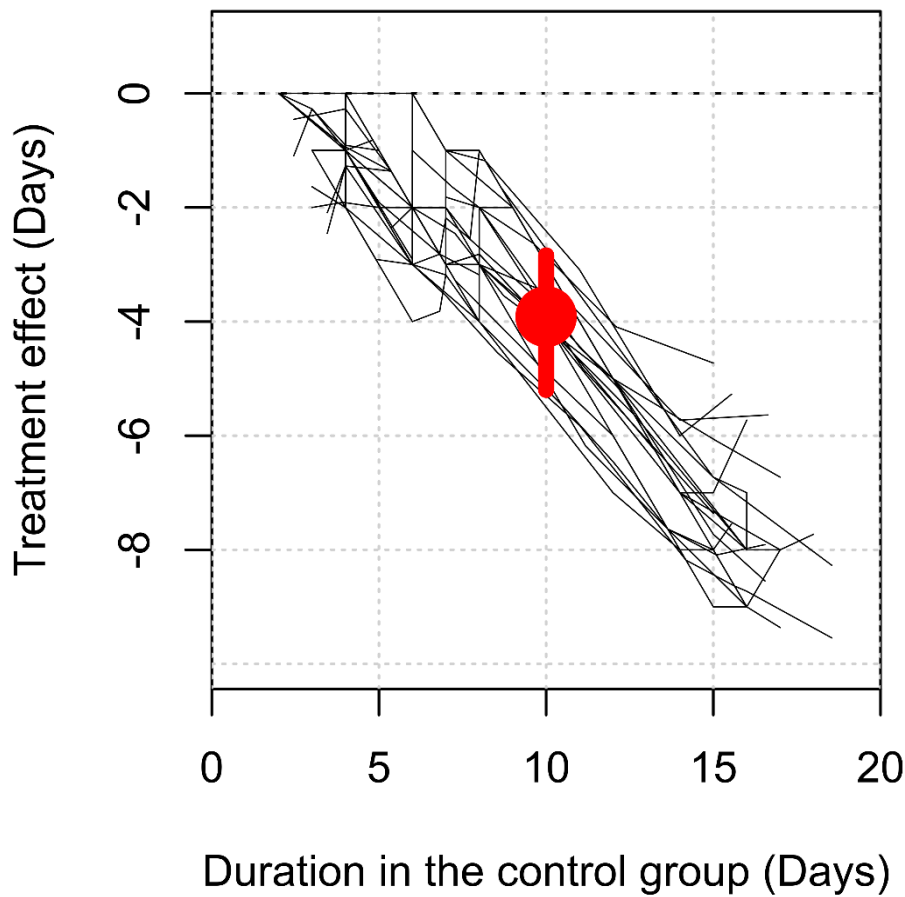

**Fig. 2.** Back-transformed quantile treatment effect (BQTE) of zinc gluconate lozenges on common cold duration in the Mossad trial in days (A) and on the relative scale (B). The dotted horizontal lines indicate the overall ATE of 4.0 days [6] in Fig. 2A, and the overall relative effect of 43% [6] in Fig. 2B. The red dots indicate the estimated BQTEs, and the vertical lines indicate the 95% CI based on bootstrap sampling. The trial analyzed: Mossad et al. [13].

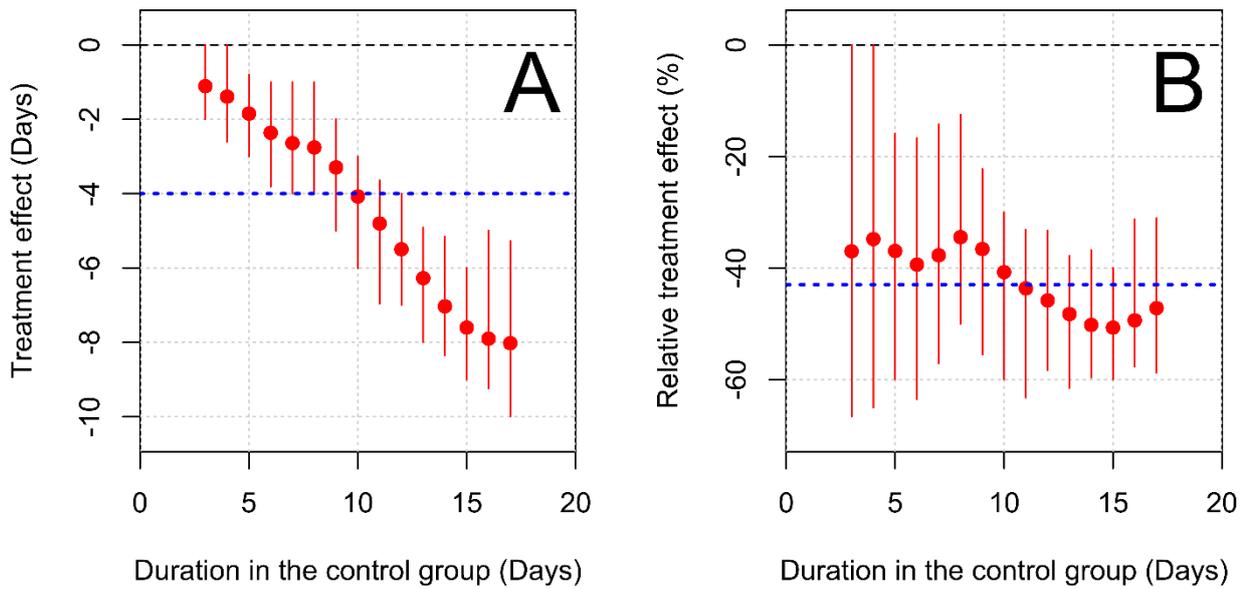

**Fig. 3.** Pooled back-transformed quantile treatment effect (BQTE) of zinc acetate lozenges on common cold duration in three randomized trials in days (A) and on the relative scale (B). The dotted horizontal lines indicate the overall ATE of 2.7 days [20] in Fig. 3A, and the overall relative effect of 36% [20] in Fig. 3B. See explanations in Fig. 2. The trials included: Petrus et al. [14], Prasad et al. [15], and Prasad et al. [16].

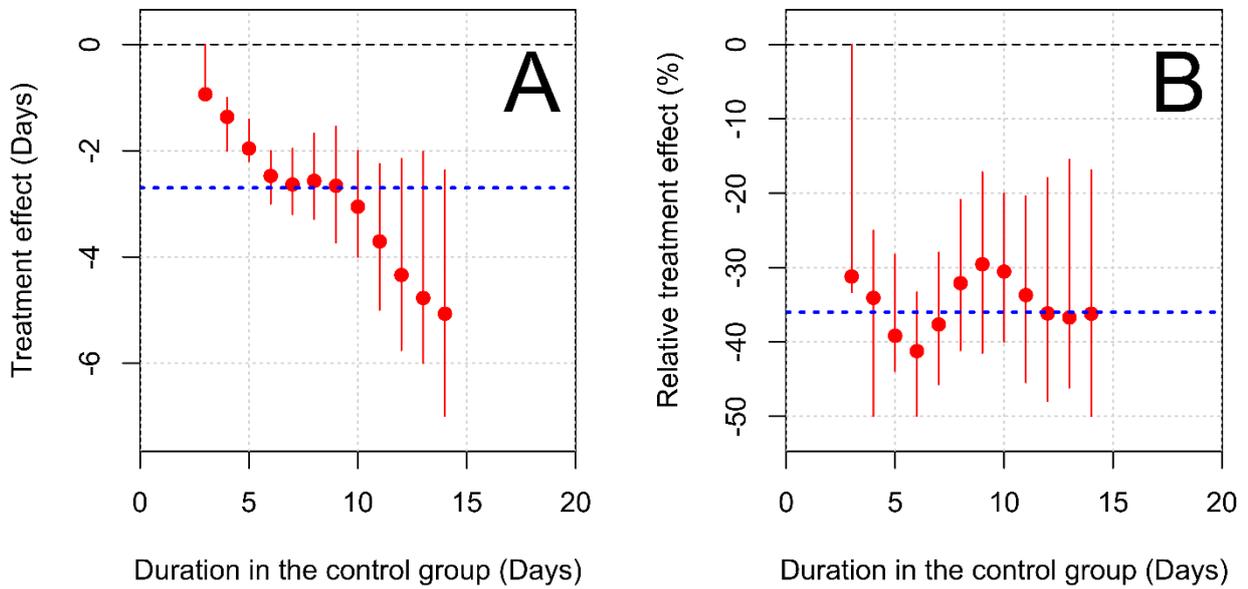

**Fig. 4.** Pooled back-transformed quantile treatment effect (BQTE) of nasal iota-carrageenan on common cold duration in two randomized trials in days (A) and on the relative scale (B). The dotted horizontal line indicates the estimated overall ATE of 1.9 days [22] in Fig. 4A; the overall relative effect was not estimated [22] and is not shown in Fig. 4B. See explanations in Fig. 2. The trials included: Ludvig et al. [17] and Fazekas et al. [18].

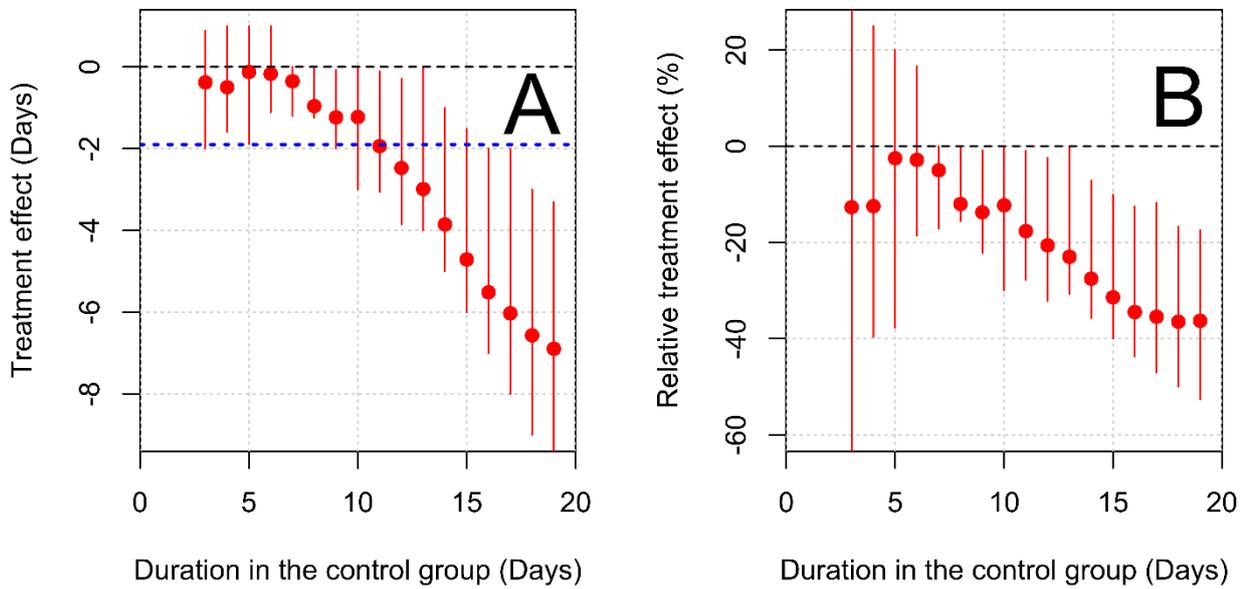